\documentclass{emulateapj} 

\def\lsim{~\raise0.3ex\hbox{$<$}\kern-0.75em{\lower0.65ex\hbox{$\sim$}}~}
\def\gsim{~\raise0.3ex\hbox{$>$}\kern-0.75em{\lower0.65ex\hbox{$\sim$}}~}
\def\lt{~\hbox{$<$}~}
\def\gt{~\hbox{$>$}~}
\def\msun{\rm\,M_\odot}
\def\lbrack2{[\![}
\def\rbrack2{]\!]}

\def\kms{\rm\,km\,s^{-1}}
\def\yr{\rm\,{\rm\,yr}}
\def\yrs{\rm\,{\rm\,yrs}}
\def\myr{\rm\,{\rm\,Myr}}
\def\myrs{\rm\,{\rm\,Myrs}}
\def\cm{\rm\,cm}
\def\erg{\rm\,erg}
\def\s{\rm\,s}
\def\pc{\rm\,pc}
\def\kpc{\rm\,kpc}

\def\mhalo{M_{\rm halo}}

\def\dt{\Delta t}
\def\mmin{M_{\rm *,min}}
\def\tdyn{t_{\rm dyn}}
\def\rhosf{\rho_{\rm SF}}
\def\dx{\Delta x}
\def\esf{\epsilon_{\rm SF}}
\def\v90{v_{90}}
\def\vmed{\langle v_{90}\rangle}

\shorttitle{DLA kinematics and outflows}
\shortauthors{Razoumov et al.}

\begin{document}
\title{DLA kinematics and outflows from starburst galaxies.}

\author{Alexei O. Razoumov\altaffilmark{1}}
\email{razoumov@ap.smu.ca}

\altaffiltext{1}{Institute for Computational Astrophysics, Dept. of
  Astronomy \& Physics, Saint Mary's University, Halifax, NS, B3H 3C3,
  Canada}


\begin{abstract}
  We present results from a numerical study of the multiphase
  interstellar medium in sub-Lyman-break galaxy protogalactic
  clumps. Such clumps are abundant at $z=3$ and are thought to be a
  major contributor to damped Ly$\alpha$ absorption. We model the
  formation of winds from these clumps and show that during star
  formation episodes they feature outflows with {\it neutral} gas
  velocity widths up to several hundred $\kms$. Such outflows are
  consistent with the observed high-velocity dispersion in DLAs. In
  our models thermal energy feedback from winds and supernovae results
  in efficient outflows only when cold ($\lsim300K$), dense
  ($\gsim100\msun\pc^{-3}$) clouds are resolved at grid resolution of
  $12\pc$. At lower $24\pc$ resolution the first signs of the
  multiphase medium are spotted; however, at this low resolution
  thermal injection of feedback energy cannot yet create hot expanding
  bubbles around star-forming regions -- instead feedback tends to
  erase high-density peaks and suppress star formation. At $12\pc$
  resolution feedback compresses cold clouds, often without disrupting
  the ongoing star formation; at the same time a larger fraction of
  feedback energy is channeled into low-density bubbles and
  winds. These winds often entrain compact neutral clumps which
  produce multi-component metal absorption lines.
\end{abstract}


\section{introduction}

Current numerical galaxy formation models can successfully reproduce
some of the properties of damped Ly$\alpha$ absorbers (DLAs), such as
the lower end ($N_{\rm HI}\le10^{21.5}\cm^{-2}$) of the column density
distribution and the total incidence rate
\citep{pontzen........08,razoumov.....08}, the distribution of metals,
and the slope of the relation between metallicity and low-ion velocity
width which appears to originate in the mass-metallicity relation in
the models \citep{pontzen........08}. On the other hand, simulations
tend to overpredict the number of DLAs with $N_{\rm
  HI}\gsim10^{21.5}\cm^{-2}$ and systematically produce fewer
high-velocity systems. Most such systems feature multiple components
in their absorption line profiles, but unfortunately one cannot map
these components from the velocity space to real space to identify the
absorption regions and constrain the mechanism producing such high
velocities.

In general, the velocity dispersion of neutral gas clouds can come
either from the gravitational infall in the process of hierarchical
buildup of galaxies, in the form of random velocities of protogalactic
clumps \citep{haehnelt..98}, or from feedback from stellar winds and
supernovae (SNe) \citep{schaye01}. In fairly massive $10^{12}\msun$
halos at $z=3$ as much as $20-30\%$ of gas by mass can be in the cold
phase surviving the infall \citep{razoumov.....08}. The corresponding
$v_{\rm circ}\sim250\kms$ can account for part of the observed neutral
gas velocity dispersion. However, more massive halos are rare at
$z=3$, and the fraction of cold gas drops sharply in $\gt10^{12}\msun$
halos, leaving us in search of other mechanisms to produce high
velocities.

Galactic winds driven by the feedback energy from stellar winds and
SNe are an obvious candidate \citep{schaye01}. Star-forming
Lyman-break galaxies (LBGs) at $z\sim3$ show evidence for large-scale
outflows with typical velocities of hundreds $\kms$
\citep{pettini.....98,pettini.......01}. In fact, with a simple
semi-analytical model \citet{mcdonald.99} showed that feedback at the
rate $1.8\times10^{50}\erg\yr^{-1}$ per $10^{12}\msun$ of halo dark
matter mass added to the velocity dispersion of neutral clouds inside
virialized halos works out perfectly to explain the observed DLA
kinematics. However, this energy transfer takes place on pc scales
currently inaccessible to cosmological models. Moreover, the inability
of numerical galaxy formation models to capture physics on such small
scales has led to a number of predicaments, the most famous of which
is the overcooling problem, accompanied by the excessive loss of
angular momentum in simulated galactic disks.







This classical problem \citep{katz92} has been somewhat alleviated in
recent years \citep{thacker.00,sommer-larsen..03} as it was realized
that feedback from young stars can be very efficient at keeping gas in
a diluted state preventing it from rapid collapse and conversion into
stars. However, even galaxy models at a sub-kpc ($0.1-1\kpc$)
resolution cannot capture propagation of supernova blastwaves into the
interstellar medium (ISM), as the injected thermal energy is radiated
away very quickly before it can be converted into kinetic energy. The
reason is very simple: the mass of a resolution element to which the
feedback energy is supplied is usually several orders of magnitude
larger than the typical mass of a SN ejecta. Therefore, the
temperature and expansion velocity of the post-shock regions are
greatly underestimated, and so is the cooling time which scales as
$\propto T^{1/2}$ above $10^7{\rm\,K}$ \citep{dallaVecchia.08}.

Cosmological simulations must then turn to ad-hoc assumptions about
the role of stellar feedback at scales below their resolution
limit. Two types of solutions have been popular. The first one is
suppressing radiative cooling in the feedback regions for the duration
of the starburst \citep{mori...97, thacker.00, sommer-larsen..03,
  stinson.....06}, to allow more efficient conversion of feedback
energy into hydrodynamical expansion. This approach leads to more
realistic simulated galaxies that correctly reproduce many of the
observed properties of present-day disk galaxies and have only with a
small deficiency in angular momentum. On the other hand due to lack of
resolution the feedback energy is supplied to a fairly large mass, and
the outflow velocities are usually underestimated. This method is
known to produce ``puffy'' galaxies that cannot reproduce the high-end
tail of the observed DLA velocity dispersion
\citep{razoumov.....08}. Since these galaxies extend to larger radii,
their outer regions may pick part of the velocity dispersion of the
local galaxy group, so that they have slightly less severe kinematics
problem \citep{pontzen........08} than similar-resolution models which
do not suppress radiative cooling.

The second widespread approach is to use kinetic feedback instead of
thermal feedback \citep{navarro.93,springel.03,dallaVecchia.08},
usually implemented in particle-based simulations. Although there are
several variations of this method, the basic idea is to give a
velocity kick to a small fraction of gas particles near the
star-forming regions, adjusting the mass loading and velocities to
reproduce observations. Some of the recipes turn off hydrodynamical
interaction of the wind particles with the gas to obtain large-scale
outflows, while other stress the importance of such interaction to
create hot bubbles in the ISM and develop galactic fountains
\citep{dallaVecchia.08}.

A popular method to circumvent some of the resolution problems that
can be combined with either of the above two approaches is to use a
sub-resolution multiphase model that describes analytically growth of
cold clouds embedded in a hot intercloud medium, star formation (SF)
in these clouds, feedback and cloud evaporation
\citep{yepes...97,springel.03}. In such models SF and feedback are
self-regulating. However, different phases are not dynamically
separated from each other, and therefore by itself such model cannot
result in outflows.






At the other end of the resolution spectrum, detailed models of small
patches of galactic disks, usually in the context of the Milky Way
galaxy, provide sufficient resolution to study turbulent ISM stirred
by SN explosions. Such models resolve hot bubbles driven by individual
SNe, fragmentation of shells created by these bubbles, and the
structure of cold dense clouds on pc scales
\citep[e.g.,][]{joung.06}. High SF rates in such models naturally lead
to galactic outflows, galactic fountains rising to several kpc away
from the midplane, and shell fragments raining back onto the disk as
intermediate-velocity cold clouds \citep{joung.06}. Such
high-resolution simulations can in principle be used to develop
subgrid models of stellar energy feedback for a given SF rate in
cosmological simulations, although to the best of our knowledge
currently there are no subgrid models in the literature that separate
the dynamics of different components of the ISM.




In the past few years it has become possible to extend such
high-resolution 3D models to entire galactic disks, albeit at a lower
spatial resolution. \citet{tasker.06a} used adaptive mesh refinement
(AMR) models to study the multiphase ISM in a quiescent Milky
Way-sized disk galaxy. They employ two SF prescriptions, one based on
cosmological simulations, with low SF threshold ($0.02\cm^{-3}$) and
low efficiency, and the other one with a much higher threshold
($10^3\cm^{-3}$) and a high efficiency. Their highest numerical
resolution is $25\pc$ which is a typical size of giant molecular
clouds; they include cooling to $300{\rm\,K}$ and later add
photoelectric heating \citep{tasker.08}. Their models reproduce a
multiphase ISM with most of the mass in cold, dense clouds, while SN
feedback drives gas out of the plane of the galaxy, but most of it
eventually falls back on the disk. All of their models reproduce the
slope of the observed relation between the SF rate and the gas surface
density, on both global and local scales, although the high-density
threshold models tend to produce more intermittent outflows and
occasionally triggered SF in the outer disk.

\citet{saitoh.......08a} carried out SPH simulations of an isolated
gas disk with $10^6-10^7$ particles to study the effect of various SF
prescriptions on the structure of the ISM. Similar to
\citet{tasker.06a,tasker.08}, they test both a cosmological
($0.1\cm^{-3}$) and a high-density ($100\cm^{-3}$) SF thresholds, but
also vary the SF efficiency $\esf$. Only the high-density threshold
models could reproduce the complex multiphase structure of the gas
disk, regardless of the value of $\esf$. In these runs the SF rates
depend on the modeled global gas flow from intermediate densities to
the actual sites of SF rather then the actual prescribed SF
efficiency. On the other hand, the low-density threshold models
produce thicker and smoother disks with SF rates highly sensitive to
the chosen value of $\esf$. Therefore, they conclude, the use of a
high SF threshold will avoid uncertainties in the SF models.







\citet{ceverino.08} developed a realistic prescription for modeling
feedback formulating conditions under which simulations would resolve
the formation of hot bubbles in the multiphase ISM. Such bubble can
only be created if simulations resolve cold dense clouds in which
feedback occurs -- only in these clouds heating can exceed cooling so
that a large fraction of feedback energy is converted into
hydrodynamical expansion, creating the hot phase and driving
winds. \citet{ceverino.08} also added heating by massive binary
systems which are ejected from molecular clouds when one of the
components becomes a SN. These ``runaway stars'' carry energy very
efficiently away from high-density regions, eventually exploding as
SNe in lower density environments and thus facilitating the formation
of the hot gas component even in low-resolution cosmological models.

In this paper the approach of \citet{ceverino.08} is used to study the
formation of winds in high-redshift protogalactic clumps responsible
for damped Ly$\alpha$ absorption. We show that a brief episode of SF
in a sub-LBG galaxy that creates a multiphase medium can also drive
winds with {\it neutral} gas velocity dispersions up to several
hundred $\kms$. If a substantial fraction of $z=3$ protogalactic
clumps form such winds at any given time, these winds can explain the
observed DLA kinematics \citep{schaye01}. Our models resolve the
effect of massive stars in protogalactic clumps. Although the spatial
resolution of this study ($12\pc$) is not sufficient to follow the
details of shell fragmentation or turbulent interactions, it is argued
that this resolution is adequate for modeling the multiphase medium
for the purposes of computing galaxy formation and launching galactic
winds in the cosmological context.







\section{Models}
\subsection{Peak cross-section of DLA absorption}

The mass range of halos that are the main contributors to the total
DLA line density is still debated. \citet{pontzen........08} argue
that the main contribution comes from halos in the mass range
$10^9-10^{11}\msun$. Lower-mass halos have much smaller absorption
cross-sections due to heating by the ultraviolet background (UVB),
while the number of halos with masses above $10^{11}\msun$ drops
sharply. \citet{pontzen........08} point out that their peak at
$\sim10^{10}\msun$ is probably related to their particular feedback
implementation in which cooling is turned off to reproduce the
blastwave solution. On the other hand, \citet{nagamine...07} see a
peak of DLA absorption shift to higher halo masses with the increased
wind strength, reaching $\sim10^{12}\msun$ in the ``strong wind''
model.

\begin{figure}
  \epsscale{1.1}
  \plotone{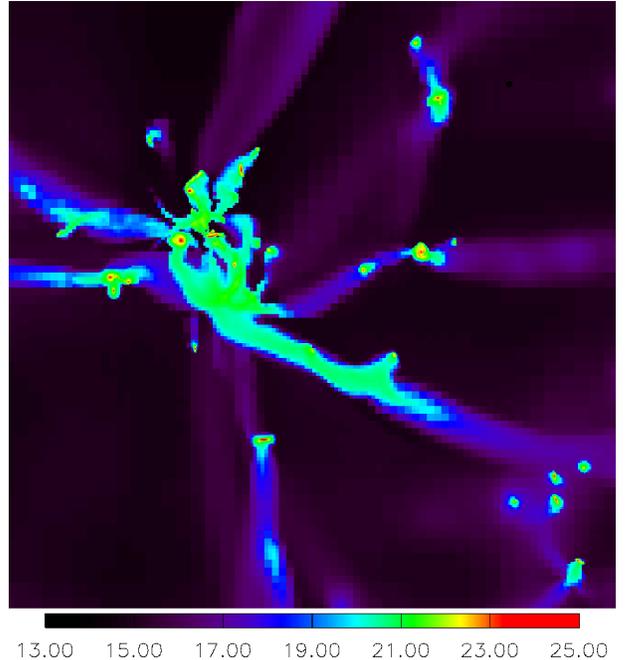}
  \caption{HI column density map of the most massive halo in model M1
    in \citet{razoumov.....08} at $z=3$ (see text for details). The
    projection is approximately $100\kpc$ on a side.}
  \label{m1-mosaic}
\end{figure}

In our earlier cosmological DLA models \citep{razoumov.....08}, the
most common DLA absorbers are halos in the range
$10^{10}-10^{11.5}\msun$. At the lower end of this range, absorption
is typically dominated by a single galaxy in the halo, while in halos
with masses above $\sim10^{11}\msun$ DLAs are commonly associated with
one of several protogalactic clumps with the average gas clump mass of
$\sim(1-2)\times10^9\msun$ (Fig.~\ref{m1-mosaic}). Our goal is to
model formation of winds in one of these clumps as it undergoes a
starburst. For the sake of simplicity, we adopt a fixed gravitational
potential with $M_{\rm halo}=3\times10^{11}\msun$ and a low $f_{\rm
  disk}=M_{\rm disk}/M_{\rm halo}=0.005$, but we argue that our
results are equally applicable to lower-mass ($10^{10}\msun$) halos
with $f_{\rm disk}\approx0.05-0.10\%$.

\subsection{Initial setup and grids}

All simulations in this paper were performed using the AMR
hydrodynamical code ENZO \citep{oshea......04}. The computational
domain is a 3D periodic box $100\kpc$ on a side covered with a $64^3$
root grid and up to seven levels of refinement corresponding to
$12\pc$ spatial resolution. A fixed spherical Navarro-Frenk-White DM
profile

\begin{equation}
M_{\rm DM}(r)=\mhalo{\ln(1+x)-{x/(1+x)}\over\ln(1+c)-{c/(1+c)}}
\end{equation}

\noindent
is assumed at the center of the volume, where $x=cr/r_{\rm vir}$, the
concentration parameter $c=12$, and $\mhalo=3\times10^{11}\msun$. The
initial distribution of gas follows an isothermal disk with a
temperature of $10^4\,{\rm K}$ and a density

\begin{equation}
\rho(r,z)=\rho_0e^{-r/r_0}{\rm sech}^2\left(z\over2z_0\right)
\end{equation}

\noindent
\citep{tasker.06a}. For all models we take $z_0=50\pc$ and
$r_0=800\pc$ to approximate a compact star-forming disk at $z=3$. The
total mass of the gas disk is then

\begin{equation}
M_{\rm disk}=f_{\rm disk}M_{\rm halo}=8\pi\rho_0r_0^2z_0.
\end{equation}

Since our disks are at least 30 times smaller than the side of the
box, using a small $100\kpc$ volume is a reasonable approximation; in
addition, small perturbations from nearby clumps should be always
expected. To resolve the initial disk configuration, inside the
central $(20\kpc)^3$ region a hierarchy of six centered nested grids
is set, with the maximum initial resolution of $24\pc$. We start with
the uniform temperature $T=10^4{\rm\,K}$, resolving the Jeans length
everywhere in the disk. During evolution we refine adaptively by the
local Jeans length requiring that it must be resolved by at least 16
cells at all times, which is four times better than the Truelove
criterion \citep{truelove.....97}.



A cooling function in the temperature range $T=300-10^9\,\rm K$ is
assumed, with heating by the ionizing UVB from \citet{razoumov...06}
with self-shielding above $0.01\cm^{-3}$. Since DLAs at $z=3$ have
fairly high metallicities at times exceeding solar, for simplicity
solar metallicity is assumed throughout all calculations.




\subsection{Star formation and feedback}

Star formation is modeled with discrete stellar particles that
represent a population of stars born in the same cell roughly at the
same time and assumed to have the same velocity vector in later
evolution. In two of the three runs presented in this paper, we adopt
the minimum stellar particle mass $\mmin=100\msun$; our stellar
particles can have any mass above $\mmin$ if a sufficient amount of
gas in a given cell satisfies the following SF criteria. A stellar
particle is always created at the finest local level of refinement, in
cells in which (1) the total gas density exceeds the threshold
$\rhosf$, and (2) the mass of the gas is larger than the local Jeans
mass. In other words, stars will be formed only in cells in which the
local Jeans length is unresolved which with our refinement criterion
is possible only at the highest AMR level. If a cell is marked as a
candidate for SF, we compute the mass of stars it would form with the
given efficiency $\esf$ over the local dynamical time $\tdyn$, and
scale that mass to the local timestep $\dt$. If
$\rho(\dx)^3\esf\dt/\tdyn$ exceeds $\mmin$, a stellar particle is
created, and the corresponding mass is removed from the gas
component. Since our minimum stellar particle mass is very low and
would allow us to record individual core-collapse SN events, we adopt
instantaneous conversion of gas into stars with $\esf$, unlike, e.g.,
in \citet{tasker.08} where the actual SF and feedback associated with
each stellar particle are continuous over the dynamical timescale.


Over its lifetime every stellar particle injects feedback energy into
the thermal energy of the gas. We use a prescription similar to that
of \citet{ceverino.08} to include feedback by both stellar winds and
type II SNe. Stellar winds supply energy at a constant rate of
$3.88\times10^{33}\erg\s^{-1}\msun^{-1}$ for $\tau_{\rm th}=40\myrs$
after creation of the stellar particle which corresponds to conversion
of $\eta_{\rm en}=2.72\times10^{-6}$ of the rest-mass energy of newly
created stars into feedback energy. In addition, during the last 10\%
of the $40\myr$ feedback phase, SNe contribute $10^{51}\erg$ per every
$100\msun$ of the initial stellar particle mass. This energy is added
to the thermal energy of the cell hosting each stellar particle, in
discrete $10^{51}\erg$ events spread uniformly over the $4\myrs$
interval. In our starburst model most stellar particle masses fall
into the range $100-200\msun$, therefore our simulations begin to
resolve environmental effect of individual massive stars exploding as
SNe.

In addition to energy, winds and SNe also return mass and metals to
the ISM. In our models, energy and mass release into the ISM is
strictly synchronized to avoid putting too much energy into regions
which have been cleared by winds and/or earlier SNe. We assume that
$\eta_{\rm mass}=0.25$ of the total mass that goes into stars is
ejected back into the ISM via winds and SNe. It can be shown that the
maximum sound speed in hot bubbles in simulations in which mass and
energy are simultaneously released into the same cell cannot exceed

\begin{equation}
c_{\rm s}\lsim\left(\eta_{\rm en}\over\eta_{\rm
    mass}\right)^{1/2}c\approx1800\kms,
\label{cspeed}
\end{equation}

\noindent
where $c$ is the speed of light, corresponding to the maximum
temperature of $\sim4\times10^8{\rm\,K}$. Actual temperatures in hot
bubbles are somewhat lower in the range $10^6-10^8{\rm\,K}$, largely
due to expansion and the work performed to compress the ambient
medium, and to a much lesser degree due to cooling in the bubble
itself. With outflow velocities added to $c_{\rm s}$, the
Courant-Friedrichs-Lewy (CFL) condition sets the shortest timesteps in
our models to several thousand years.








In all our runs the SF efficiency $\esf=0.3$ is assumed. Many authors
have found that the exact value of $\esf$ has little impact on the
mean SF rates \citep[e.g.]{stinson.....06}, as long as it is in the
range from 0.05 to 1 (see their Fig.14). Moreover, the true efficiency
of SF, i.e. the fraction of gas that is eventually converted into
stars in dense clouds should be determined by an interplay of various
processes starting from the hydrodynamical timescale of gas supply to
the star-forming regions. \citet{saitoh.......08a} have found that at
sufficiently high $\rhosf$ the SF rates are effectively set by the
timescale of cold gas supply from reservoirs ($n_{\rm H}=1\cm^{-3}$)
to the star-forming regions ($n_{\rm H}\gsim100\cm^{-3}$) which in
their calculations is about five times longer than the local dynamical
timescale in the star-forming regions.

In this paper we are using a set of three simulations listed in Table
1: a high-resolution starburst model A1, a high-resolution quiescent
disk model A2, and a low-resolution model A3 for which $\rhosf$ was
adjusted to produce the highest SF rate. The high-resolution models
used 7 levels of AMR in the $(10\kpc)^2\times6\kpc$ region centered on
the disk resulting in $12\pc$ grid resolution. The low-resolution
model employed 6 levels of refinement in the same region corresponding
to $24\pc$ spatial resolution.


We ran the quiescent model to estimate the effect of star formation on
the structure of the ISM and on galactic wind kinematics. There are
two ways to reduce the SF rate in our models: increase the SF density
threshold $\rhosf$ or increase the minimum stellar particle mass
$\mmin$. Note that in our setup $\mmin$ is not independent of
$\rhosf$. A star particle is formed only if $\rho(\dx)^3\esf\dt/\tdyn$
in a cell exceeds $\mmin$. For the fiducial value
$\rhosf=158\msun\pc^{-3}$ and $\esf=0.3$, the gas mass in a cell
allowed to form stars is

\begin{equation}
M_{*,{\rm cell}}=270\msun\left(\rho\over158\msun\pc^{-3}\right)\left(\dt\over2000\yrs\right).
\end{equation}

\noindent
Setting $\mmin=100\msun$ would result in immediate SF once the density
exceeds $\rhosf$, whereas using a much higher value would delay SF
until more gas accumulates in the cell. For the quiescent disk model
we use $\mmin=1000\msun$. This prescription ultimately results in
conversion of approximately the same amount of gas into stars, but
over a several times longer period, and produces a very different ISM
morphology and much weaker winds.







\begin{table}
  \begin{center}
    \caption{Simulation parameters: (1) highest level of refinement
      $l_{\rm max}$, (2) SF threshold $\rhosf$, (3) minimum stellar
      particle mass $\mmin$.
      \label{modelList}}
    \begin{tabular}{cccc}
      \hline\hline
      model & $l_{\rm max}$ & $\rhosf/(\msun\pc^{-3})$ & $\mmin/\msun$\\
      \hline
      \hline
      A1 & 7 & 158 & 100\\ 
      A2 & 7 & 158 & 1000\\ 
      A3 & 6 &  25 & 100\\ 
      \hline
    \end{tabular}
  \end{center}
\end{table}









\section{Results}

\subsection{Global ISM morphology}

\begin{figure}
  \epsscale{1.2}
  \plotone{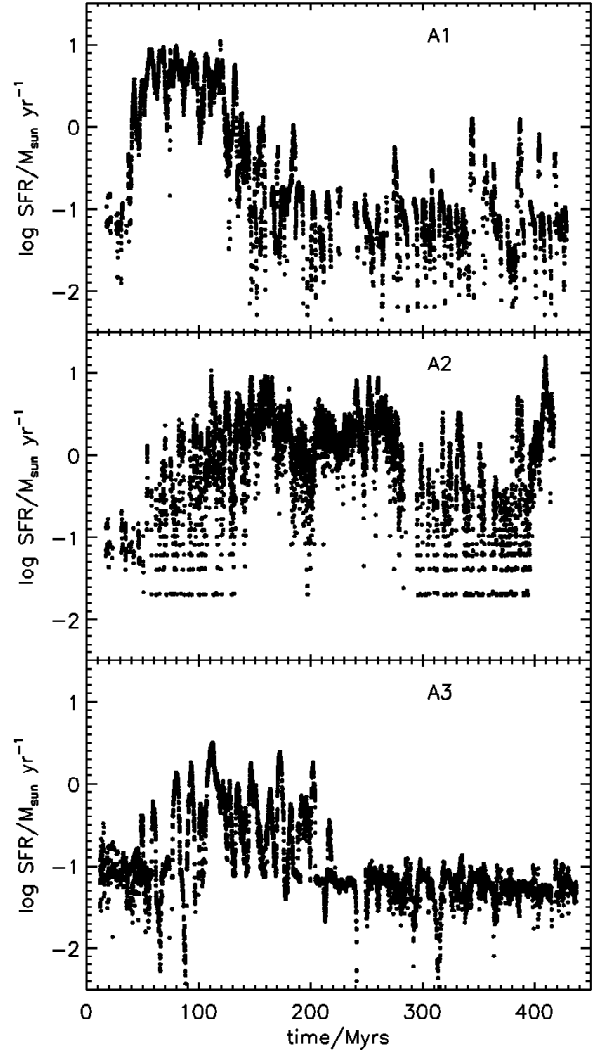}
  \caption{Total SF rates in all three runs sampled at 50 kyr time
    intervals.}
  \label{sfr}
\end{figure}

Without cooling our disks would be marginally Toomre-unstable. Adding
cooling leads to rapid gas accumulation near the galactic midplane and
its subsequent fragmentation into cold clumps and warm interclump
material. For the initial central disk density $10^{-22}{\rm g}\cm^3$
and temperature $10^4{\rm\,K}$ the cooling time is of order of few
years leading to gas collapse of the disk onto the midplane on the
timescale of $\sim20\myrs$. Soon thereafter first cold clumps form in
which SF begins. By $50\myrs$ high-resolution models start developing
a complex ISM morphology characterized by dense clouds and filaments
separated by warm ($10,000-20,000{\rm\,K}$) medium seen in many
simulations \citep[e.g.,][]{wada.01}, and to a lesser degree in the
low-resolution model. In model A1 ample gas supply quickly leads to a
starburst starting at $t\approx50\myrs$ and lasting $\sim80\myrs$
(Fig.~\ref{sfr}) in which $\sim20\%$ of the gas in the disk is
converted into stars. In the quiescent disk model A2 there is no
single starburst phase; the first significant episode of gas
conversion into stars takes place well past $t=100\myrs$, with
intermittent SF throughout the entire run. It is interesting that by
the end of simulation A2 at $t\approx420\myrs$ approximately the same
total stellar mass ($3\times10^8\msun$) is accumulated, although its
effect on the underlying gas distribution will be completely
different.

\begin{figure}
  \epsscale{1.}
  \plotone{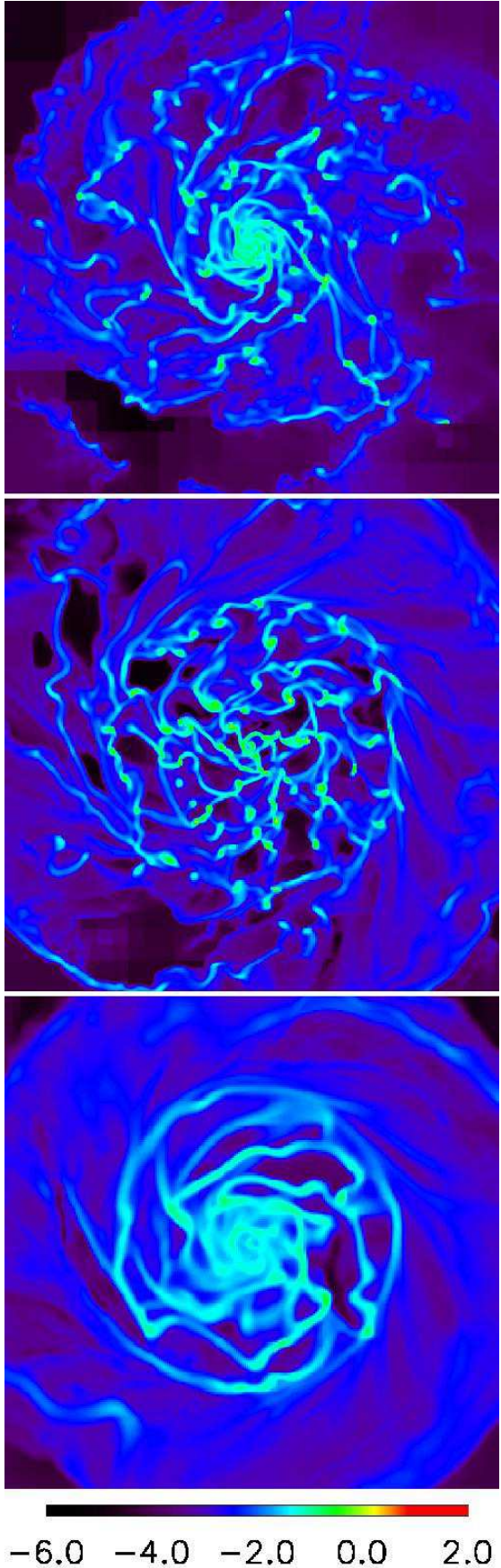}
  \caption{Disk surface density of models A1, A2, A3 at
    $t=119\myrs$.}
  \label{disks0439}
\end{figure}

\begin{figure}
  \epsscale{1.2}
  \plotone{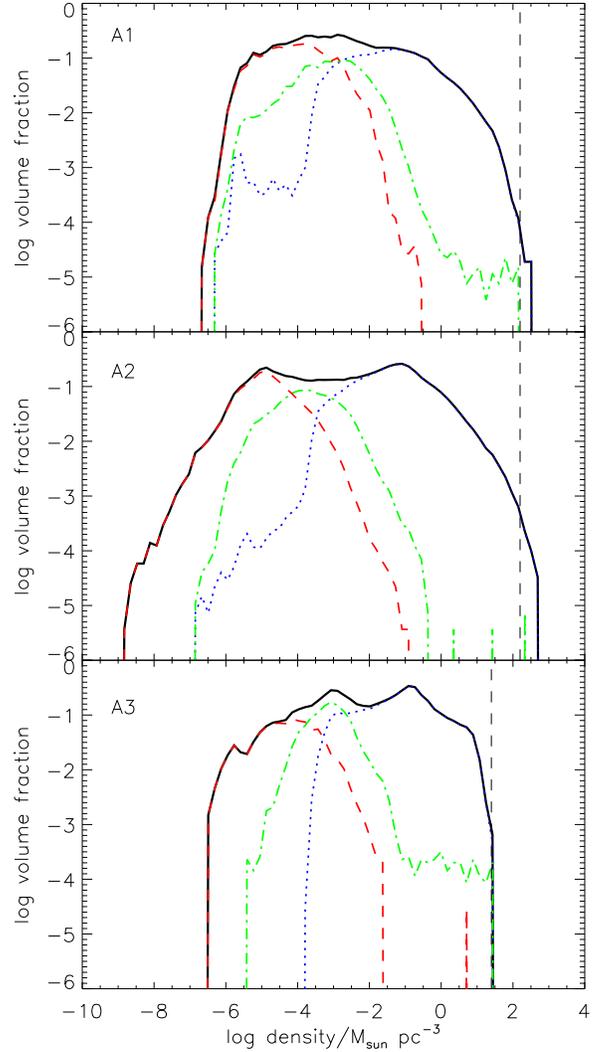}
  \caption{PDF of gas density in the disk, defined as the fraction of
    the volume per unit logarithm density interval (thick solid line),
    in all three models at $t=119\myrs$. Only cells in the
    $r\le10\kpc$, $|z|\le100\pc$ region are shown. All three
    components of the ISM -- cold ($T\lt10^3{\,\rm K}$, dotted line),
    warm ($10^3{\,\rm K}\lt T\lt10^4{\,\rm K}$, dash-dotted line), and
    hot ($T\gt10^4{\,\rm K}$, dashed line) -- are clearly
    visible. Vertical dashed lines show the SF thresholds.}
  \label{gasPDF}
\end{figure}

The morphology of the ISM is clearly affected by feedback from SF as
can be seen from the surface density maps in Fig.~\ref{disks0439} at
$t=119\myrs$ corresponding to the end of the starburst phase in model
A1. By this time in the starburst model a much larger amount of mass
and energy have been injected through feedback into the lower-density
gas. The result is a much larger role of pressure confinement of cold
clouds in model A1, as opposed to more gravitational confinement in
the quiescent model. The mass distribution appears to be smoother in
the starburst model, with a lower density contrast between the clumps
and the voids (Fig.~\ref{gasPDF}). Also evident in model A1 is a
more pronounced gas accumulation near the galactic center, and a
violent stripping of the outer gas regions of the disk by feedback
waves. The latter process depends, of course, on the gas mass of the
outer disk which in turn is determined by the cosmic accretion which
we do not compute in our current models.

In our high-resolution models dense clouds are continuously being
formed and destroyed by self-gravity, differential rotation, feedback
from SF inside the clouds, and interaction with feedback waves coming
from nearby star-forming regions. Any single cloud usually survives
only for a fraction of its galactic orbital revolution, in other
words, from few Myrs to few tens of Myrs. This is consistent with many
estimates of the giant molecular cloud (GMC) lifetimes in the Milky
Way galaxy, although we do not resolve the scales and processes taking
place inside these clouds.

\subsection{Conditions in simulated SF regions}

We are interested in modeling conditions in the star-forming regions
that facilitate launching of galactic winds from thermal feedback
only, without suppression of cooling. We will here review a set of
criteria necessary to model winds and expanding hot bubbles in the
ISM. First and foremost, heating by a SN must lead to a sharp rise in
the gas temperature that would drive the hot bubble expansion without
rapid cooling. In other words, during all stages of bubble expansion
heating must dominate over radiative cooling. For the early stages,
this condition was elegantly formulated in \cite{ceverino.08} (see
their eq.~5); using our SF threshold, we will write it as

\begin{eqnarray}
\Gamma&\gsim&7.8\times10^{38}\erg\s^{-1}\msun^{-1}\left(\rho\over158\msun\pc^{-3}\right)\times\\
&&\left(\Lambda\over10^{-22}\erg\cm^3\s^{-1}\right)\left(\rho_*\over\rho\right)^{-1},
\label{heatingDominates}
\end{eqnarray}

\noindent
where $\rho_*$ is the spatial density of young stars, and
$\rho_*/\rho$ is expected to be in the range $0.1-1$. If the gas
temperature is around $10^4{\rm\,K}$, cooling
$\Lambda\sim10^{-22}\erg\cm^3\s^{-1}$, and heating from SNe cannot
counterbalance cooling in any moderate overdensity. On the other hand,
at very low temperatures ($\sim100{\rm\,K}$) cooling is much less
efficient ($\Lambda\sim10^{-25}\erg\cm^3\s^{-1}$), and even at high
star-forming cloud densities feedback may be able to heat the
gas. Therefore, \citet{ceverino.08} argue, it is crucial to include
cooling to $\sim100{\rm\,K}$ to resolve the cold phase in order to
heat up the gas via SN feedback. Once hydrodynamical expansion of the
feedback region begins, gas flows out, the mass ratio $\rho_*/\rho$
increases, assisting further heating and expansion.

When a SN injects energy into the ISM, the resulting pressure in the
hot bubble greatly exceeds the surrounding pressure. Provided that the
energy is not quickly radiated away, the bubble expands only if it is
not confined by self-gravity. This second condition was formulated in
\citet{ceverino.08} in terms of the pressure difference between the
bubble and the surrounding gas

\begin{equation}
\Delta p\gsim{4\pi G\over3}(\rho r)^2,
\label{pressureDifference}
\end{equation}

\noindent
where $r$ is the radius of the bubble, and $\rho$ is the ambient gas
density. Using the ideal gas equation of state and our SF density
threshold, we can rewrite eq.~\ref{pressureDifference} to obtain the
minimum resolution necessary to model the expansion of the HII regions
against self-gravity

\begin{equation}
  \dx\sim2r\lsim14\pc\left(T\over10^4{\rm\,K}\right)^{1/2}
  \left(\rho\over158\msun\pc^{-3}\right)^{-1/2}.
\label{minimumResolution1}
\end{equation}


Since our models start to resolve individual SNe, the mass of a
resolution element should be small enough in order for it to get
heated by the typical $\sim10^{51}\erg$ explosion energy. A single SN
explosion in a dense cloud may have a hydrodynamical impact only if it
can heat its host cell to high ($10^6-10^8{\rm\,K}$) temperatures of
an expanding hot bubble. In other words, the energy input per SN
should then exceed

\begin{equation}
E_{\rm SN}\gsim{3\rho(\dx)^3\over2\mu m_{\rm H}}kT,
\end{equation}

\noindent
which gives us an estimate of the minimum resolution necessary to heat
up the host cell


\begin{eqnarray}
\dx&\lsim&2.49\pc\left(E_{\rm SN}\over10^{51}\erg\right)^{1/3}\\
&&\times\left(\rho\over158\msun\pc^{-3}\right)^{-1/3}
\left(T\over10^6{\rm\,K}\right)^{-1/3}.
\label{minimumResolution2}
\end{eqnarray}

\noindent
At first glance, this constraint requires much higher resolution than
the self-gravity condition
(eq.~\ref{minimumResolution1}). Fortunately, type II SN explosions
have a $30-40\myr$ delay after the initial starburst, and many of them
explode in environments which have been previously cleared by stellar
winds and neighbouring SNe. Even more importantly, the lifetimes of
individual cold clouds are usually in the range from few Myrs to few
tens of Myrs. By the time a stellar particle hosts a SN explosion, its
cloud of origin is very likely to have been destroyed, and the SN
energy is released into a $\rho\ll\rhosf$ environment. In addition,
stellar particles may have non-negligible intrinsic velocities --
traveling even a small $5\kms$ velocity for $35\myrs$ will take a
particle $180\pc$ away from its birthplace. Therefore, the ISM
densities in which type II SN explosions take place are likely to be
several orders of magnitude smaller than $\rhosf$ making the
constraint in eq.~\ref{minimumResolution2} much less demanding.


Once conversion of feedback energy into hydrodynamical expansion
becomes efficient, lack of spatial resolution can present an
additional problem. If the density contrast in the ISM is not resolved
in the simulation, heating and hydrodynamical stirring of the
star-forming clouds might erase high-density peaks bringing the
ongoing SF to a halt. In other words, SF/feedback can be too
self-regulating at low resolution. Since we do not know a priori the
amount of clumping in the ISM at $z=3$, perhaps the most reliable way
to reduce this effect is to compare the SF rates at various
resolutions.







\subsection{Neutral gas kinematics in quasar absorption lines}

\begin{figure}
  \epsscale{1.2}
  \plotone{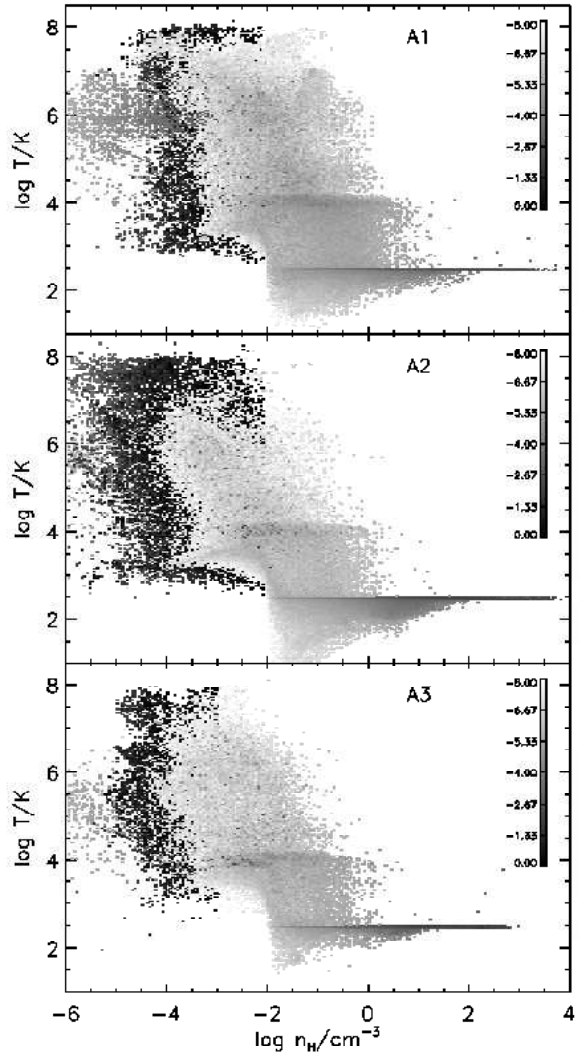}
  \caption{$\rho-T$ diagram of the entire simulation volume in each
    run at $119\myrs$. Each plot is divided into $200^2$ cells, and
    each cell is colored by its mass fraction (the darker means a
    higher mass fraction).}
  \label{rhotemp}
\end{figure}

Fig.~\ref{rhotemp} shows the $\rho-T$ diagram of the entire simulation
volume in all three runs at $t=119\myrs$. In cells hosting stellar
particles, stellar winds and SNe return both mass and energy. Assuming
that winds remove all ambient gas, the minimum density in such cells
can be easily estimated from the mass loss rate of each star particle
of mass $M_*$ during its feedback stage

\begin{eqnarray}
  \rho_{\rm low}&\approx&{0.25M_*\over40\myrs}{\Delta x\over2v_{\rm flow}}
  {1\over\Delta x^3}\approx8.6\times10^{-5}{\cm^{-3}}\\
  &&\times\left(v_{\rm flow}\over1000\kms\right)^{-1}
  \left(M_*\over100\msun\right)
  \left(\Delta x\over12\pc\right)^{-2},\nonumber
\end{eqnarray}

\noindent
where $v_{\rm flow}$ is the fiducial outflow speed, and $\Delta x$ is
the cell size. These feedback regions can be easily seen in the
starburst model in Fig.~\ref{rhotemp} at $T\sim10^6{\rm\,K}$, and to a
much lesser degree in the quiescent model.

Since we model isolated systems without external accretion, winds from
the disk do not experience any ram pressure of the infalling material
and can be stopped only by gravity and collision with the gas
previously blown off the disk. The wind speeds often exceed several
hundred $\kms$, consequently a large fraction of the volume is quickly
filled by a low-density ($\sim10^{-4}\cm^{-3}$) gas
(Fig.~\ref{rhotemp}). Since we model a large ($100\kpc$) computational
volume, the mass fraction locked in this low-density component is
significant, especially in the quiescent model A2. In the starburst
model A1 the density of the wind is clearly much higher
(Fig.~\ref{rhotemp}). Can such dense winds from starburst environments
account for the wide absorption line profiles seen in DLAs?


\begin{figure}
  \epsscale{1.2}
  \plotone{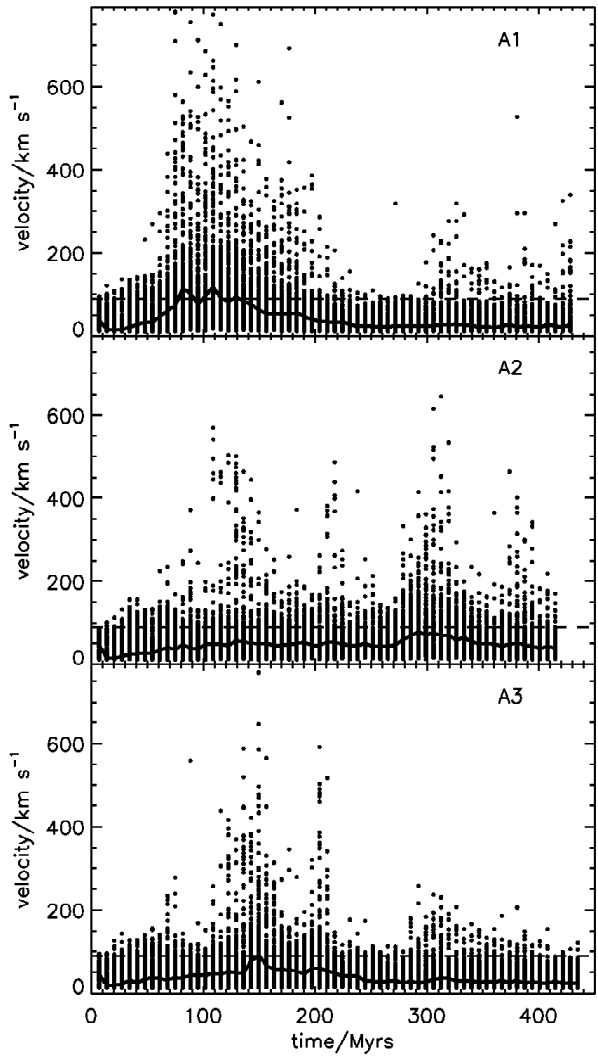}
  \caption{$\v90$ absorption velocity widths of low-ionization lines
    in three models as a function of time. The solid line in each
    panel shows the median velocity. The dashed horizontal line
    indicates the observed $90\kms$ DLA median velocity.}
  \label{velocityTime}
\end{figure}


To answer this question, we constructed a set of low-ionization metal
line spectra. At each time output, we projected 200 random lines of
sight within $3\kpc$ of the center of each disk and calculated
absorption line profiles of an unsaturated low-ion transition along
sight-lines with HI column density above $10^{20.3}\cm^{-2}$. For each
such profile we calculated a line width $\v90$ corresponding to $90\%$
of the total optical depth of all components in the line. Note that
this diagnostic measures the neutral gas velocity dispersion, not the
typical outflow velocities, and is dominated by clouds with large
optical depths – its more detailed discussion and the comparison to
the equivalent width can be found in
\citet{prochaska....08}. Fig.~\ref{velocityTime} shows the
distribution of $\v90$ widths for each model as a function of time,
along with the median value. Although these velocities cannot be
compared to DLAs statistics directly, since we do not have a
cosmological sample and do not account for gas accretion which would
regulate SF episodes, we note that the typical absorption velocities
are much higher in the starburst model. None of the cosmological DLA
models can reproduce the observed median $\vmed\approx90\kms$, with
the actual value of $\vmed$ of order $40-50\kms$ in
\citet{razoumov.....08} and close to $60\kms$ in
\citep{pontzen........08}. Note that in cosmological models the
simulated widths are also sensitive to the DLA cross-sections, as, for
instance, the velocity dispersion in ``puffy'' galaxies with extended
radial profiles will have an additional weight.

We argue that if a substantial fraction of clumps along the quasar
line of sight in a host halo experience an active starburst, it could
result in a much larger velocity dispersion possibly explaining the
observed incidence rate of high-velocity DLAs. The fraction of such
active star-forming galaxies is poorly constrained and can be only
computed from the cosmic gas infall rate onto individual systems in
cosmological simulations.

\begin{figure}
  \epsscale{1.2}
  \plotone{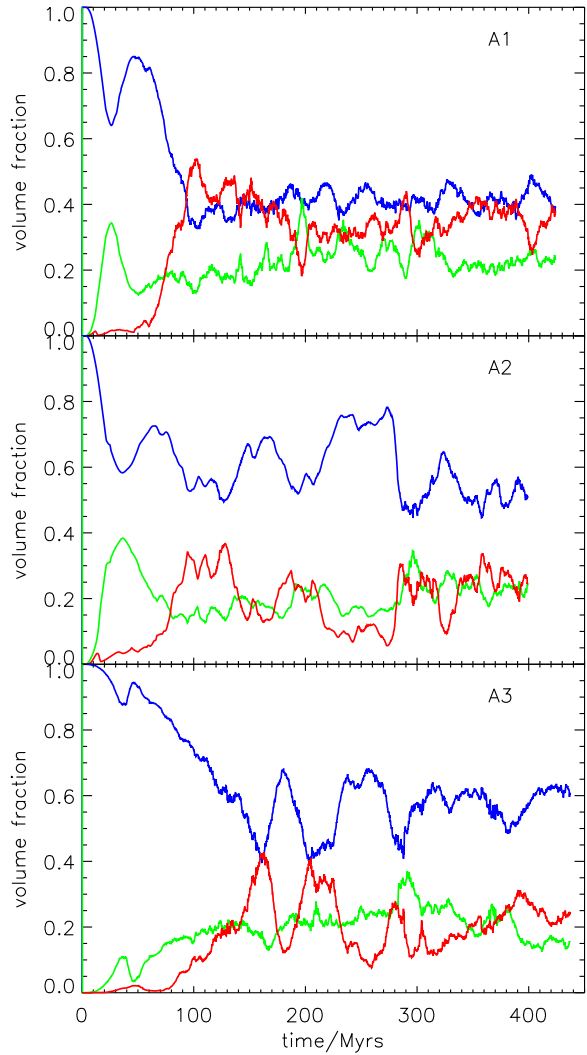}
  \caption{Volume fraction of cold ($T\lt10^3{\rm\,K}$, blue), warm
    ($10^3-3\times10^4{\rm\,K}$, green), and hot
    ($T\gt3\times10^4{\rm\,K}$, red) gas in the galactic midplane
    ($|z|\le50\pc$) for all three models.}
  \label{phases}
\end{figure}

Here instead we focus on individual systems. In the starburst model
$\vmed$ approximately matches the observed value for $60\myrs$,
whereas the quiescent model has $\vmed\approx50\kms$ for most of the
disk evolution. The low-resolution model features a delayed and much
weaker SF resulting in a brief episode during which $\vmed$ reaches
the observed value. Fig.~\ref{phases} shows the time evolution of the
volume fractions of cold, warm and hot gas in the plane of the
disk. After very rapid cooling the early ($t\lt50\myrs$) evolution is
marked by separation of gas into warm and cold components. Shortly
thereafter feedback gives rise to the hot component, and subsequent
evolution of the disk is characterized by recurrent episodes of gas
heating and cooling. The ``depth'' of these episodes is higher in the
low-resolution model, in which relatively low-density cold clumps are
more susceptible to feedback. At high resolution the relative change
of the volume fraction of cold and hot gas is visibly reduced, as the
number of cold star-forming clumps increases, and so does the density
in individual clumps. The higher SF rate results in a larger volume
filling fraction of the hot gas, at the same time driving more
energetic winds from the disk.

\begin{figure*}
  \epsscale{1.1}
  \plotone{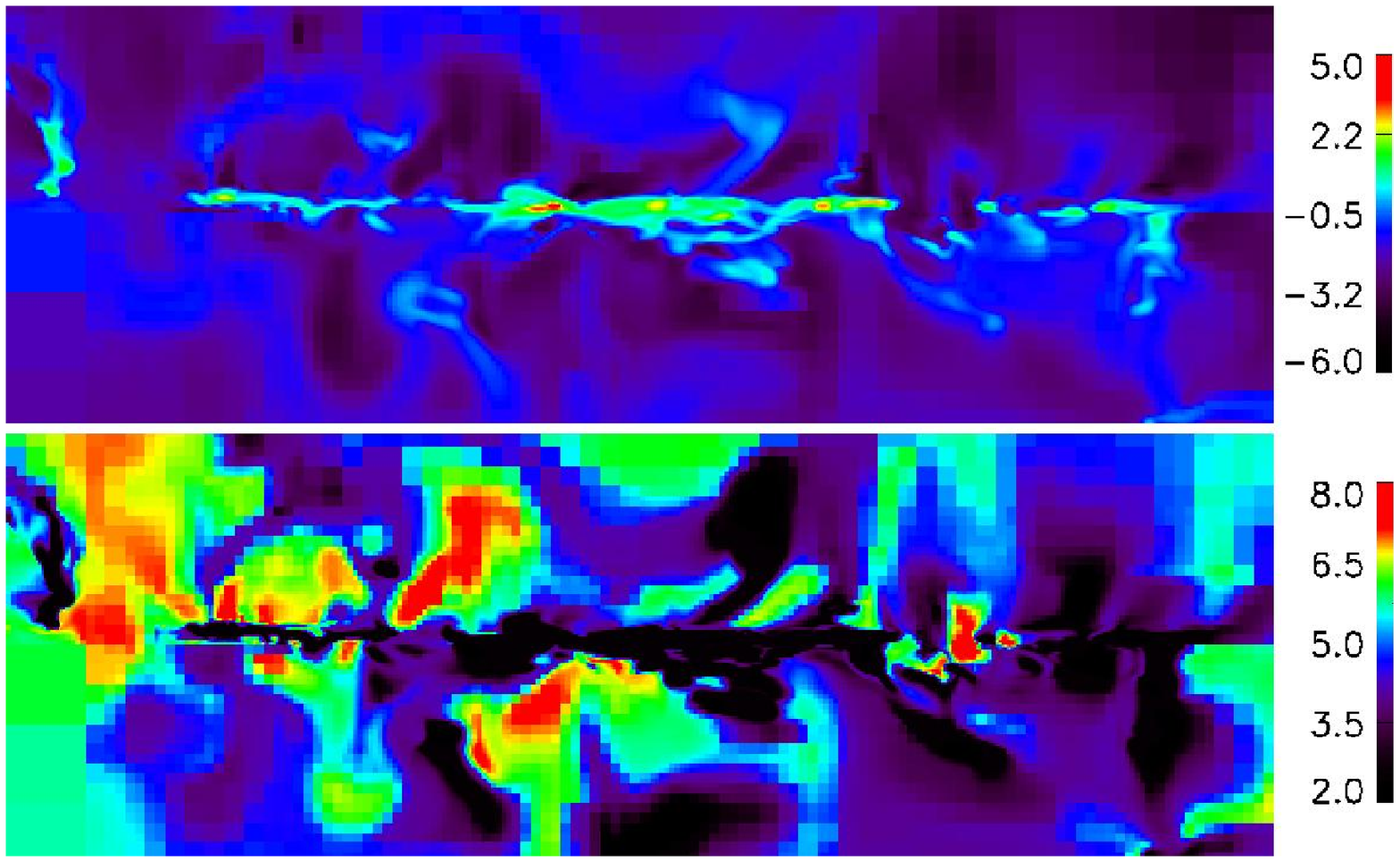}
  \caption{Edge-on slice through the disk of model A1 at $119\myrs$
    showing gas density (top panel, $\log(n_{\rm H}/\cm^{-3})$) and
    temperature (bottom panel, $\log(T/{\rm\,K})$).}
  \label{slice-sn02-0439}
\end{figure*}

We can see the formation of hot galactic chimneys driving the outflows
in the vertical slice in Fig.~\ref{slice-sn02-0439}. The visual
structure of the outflows is very different from the single-source
models of winds from high-redshift dwarf galaxies
\citep[e.g.,][]{fujita...04}, more resembling the high-resolution
multiphase models of \citet{ceverino.08,wada07}. Our current spatial
resolution is not yet sufficient to model instabilities in the shells
created by hot bubbles \citep{ferrara.06} or even follow these shells
farther away from the disk.

\begin{figure}
  \epsscale{1.2}
  \plotone{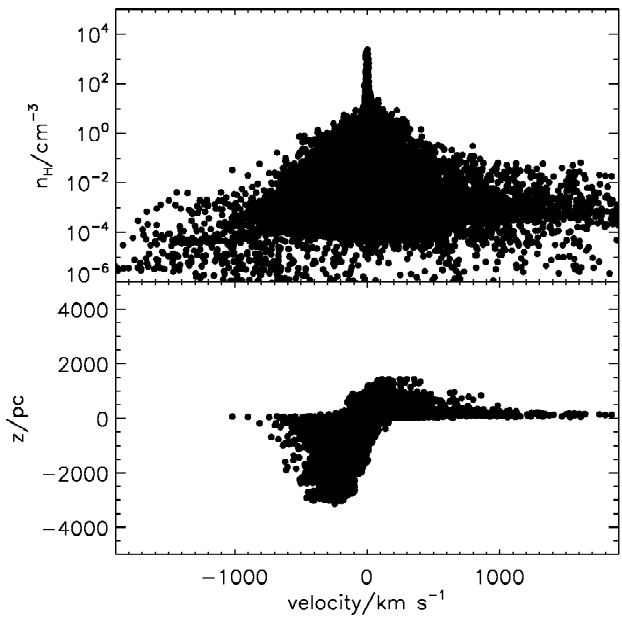}
  \caption{Outflow kinematics in model A1 at $t=119\myrs$. Top
    panel: scatter plot of gas density vs. the vertical velocity
    component. Lower panel: height above/below the disk vs. the
    vertical velocity component for all $n_{\rm H}\gt0.01\cm^{-3}$
    cells.}
  \label{velz-sn02-0439}
\end{figure}

\begin{figure}
  \epsscale{1.1}
  \plotone{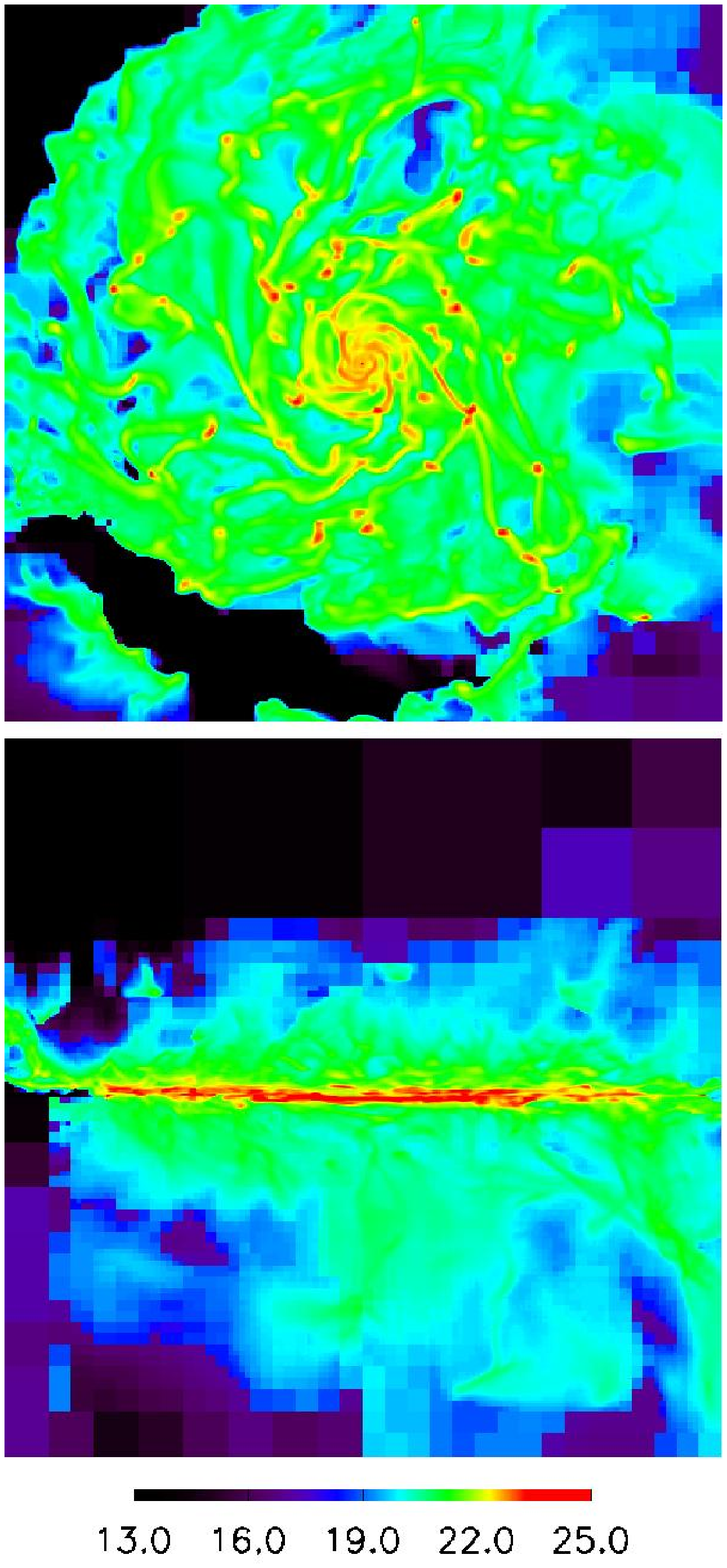}
  \caption{HI column density of model A1 at $t=119\myrs$.}
  \label{hi-sn02-0439}
\end{figure}

Fig.~\ref{velz-sn02-0439} shows the range of velocities and densities
in the wind in the starburst model. Ionized, low-density wind moves
with velocities up to several thousand $\kms$, whereas neutral gas
exhibits velocities of few hundred $\kms$. Cold gas absorption comes
from $|z|\lsim3\kpc$; there is a clear asymmetry above and below the
disk, explained by the fact that feedback starts in a fairly small
number of cold clumps. This asymmetry can be also seen in the HI
column density maps in Fig.~\ref{hi-sn02-0439}. It is important to
remember that since our models do not account for interaction of winds
and shells with the infalling material, the extent of the HI absorbing
regions might change in more realistic models with cosmic infall.

\section{Conclusions}

We use high-resolution hydrodynamic simulations of isolated $z=3$
protogalactic clumps to show for the first time that the high-end tail
of the DLA neutral gas velocity width distribution can be naturally
produced if a substantial fraction of clumps in $10^{10}-{\rm
  few}\times10^{11}\msun$ halos experience a starburst event at the
rate $\sim5\msun\yr^{-1}$ for a sustained period of few tens of
Myrs. Such starbursts would produce a multiphase ISM in which the
volume fraction of the hot gas exceeds $50\%$ in the midplane of the
clump. Hot bubbles created by feedback expand both in the horizontal
direction to fill the interclump material, and vertically to form
chimney-like structures which lead to high-velocity winds. Shells and
neutral gas fragments embedded in such winds $1-3\kpc$ above the disk
give rise to neutral gas absorption with the median velocity
$\vmed\approx90\kms$ lasting roughly for the entire duration of the
starburst.

Similar to other multiphase disk models, the cold component of the ISM
is distributed in a complex network of filamentary structures confined
by hot bubbles and voids. Inside these filaments dense clouds form as
gravitational instabilities the growth of which is assisted by the
external pressure. Once their growth begins, the clouds proceed to
form stars very quickly. Only a small fraction of each cloud
($\sim1-2\%$ by mass) is converted into stars, as the clouds are being
continuously destroyed by stellar winds from inside, interaction with
shells pushed by rapidly expanding nearby hot bubbles, and collisions
with other clouds. In our simulation clouds are transient objects
constantly exchanging mass and energy with the cold shells and
filaments, as well as with the hot gas, and rarely surviving as
discrete entities for longer than a fraction of the rotational period.


In this paper we chose the typical clump masses and initial conditions
representative of protogalactic environments at $z=3$. We expect
similar winds to arise in grid-based cosmological simulations at
$\sim10\pc$ spatial resolution. Although none of the current
cosmological models have such resolution, we argue that using AMR
techniques to zoom in only on those protogalactic clumps that have a
high gas infall rate will allow to obtain galactic winds from thermal
feedback in the cosmological context, without suppressing cooling or
using kinetic feedback.

\section*{Acknowledgments}

I thank Jesper Sommer-Larsen and Eduard Vorobyov for many useful
discussions. I am grateful to Marc Schartmann for providing me with
the cooling data. Computational facilities for this work were provided
by ACEnet, the regional high performance computing consortium for
universities in Atlantic Canada. ACEnet is funded by the Canada
Foundation for Innovation (CFI), the Atlantic Canada Opportunities
Agency (ACOA), and the provinces of Newfoundland \& Labrador, Nova
Scotia, and New Brunswick. I acknowledge financial support from
ACEnet.

\bibliographystyle{apj}


\end{document}